\documentstyle[aps,epsf,floats]{revtex}
\begin{document}
\newcommand{\eg}{\mbox{e.\ g.\ }}
\newcommand{\ie}{\mbox{i.\ e.\ }}
\twocolumn[\hsize\textwidth\columnwidth\hsize\csname
@twocolumnfalse\endcsname
\title{Frequency Scaling of Microwave Conductivity in the Integer Quantum Hall Effect Minima}
\author{R.\ M.\  Lewis,  J.\ P.\  Carini}
\address{Department of Physics, Indiana University, Bloomington, Indiana 47405 \\ }
\date{\today}

\maketitle

\begin{abstract} We measure the longitudinal conductivity $\sigma_{xx}$ at frequencies $1.246 {\rm \ GHz} \ \le f \le 10.05$ GHz over a range of temperatures $235 {\rm \ mK}  \le T \le 4.2$ K with particular emphasis on the Quantum Hall plateaus.  We find that  $Re(\sigma_{xx})$ scales linearly with frequency for a range of magnetic field around the center of the plateaus, i.e. where $\sigma_{xx}(\omega) \gg \sigma_{xx}^{DC}$.  The width of this scaling region decreases with higher temperature and vanishes by 1.2 K altogether. Comparison between localization length determined from $\sigma_{xx}(\omega)$  and DC measurements on the same wafer show good agreement.

\end{abstract}
\pacs{PACS numbers;}
\vskip2pc]
\narrowtext
\smallskip

The Integer Quantum Hall Effect (IQHE) \cite{vonklitzing,prange_girvin} offers a system in which the disorder of a 2 dimensional electron gas (2DEG) can, in effect, be tuned by adjusting the perpendicular magnetic field (B).  Measurements of the longitudinal conductivity ($\sigma_{xx}$) at finite temperature ($T$) show that for a narrow range of B corresponding to the transition between Hall plateaus the system is weakly conducting, and the localization length ($\xi$) is said to have exceeded the dephasing length $l_{\phi}$ \cite{hpwei,pruisken} and $\sigma_{xx} \sim \frac{e^2}{2h}$.  However, across the plateaus ($\sigma_{xy}=\frac{e^2}{\nu h}$, where $\nu$ is the Landau level filling factor) the IQHE is insulating, the Fermi energy is in a gap in the density of states (E$_g$), and $\xi$ shrinks to aproximately the magnetic length at the center of the plateaus.  Thus, as $ T \longrightarrow 0$, conduction in the plateaus takes the form of hopping and $\sigma_{xx} \ll \frac{e^2}{2h}$ \cite{EShopping,ESbook84}.  When Coulomb effects are important, the DC hopping conduction obeys
\begin{equation}
\sigma_{xx}(T)= \frac{\sigma_0}{ T}{\rm \ e}^{- (T_0/T)^{\frac{1}{2}}} \ {\rm where} \ T_0= \frac{6.2 \ e^2}{4 \pi \epsilon_0 \epsilon k_B \xi(\nu)}
\end{equation}
in a variety of measurements \cite{gebert,dahm}.  And, because $T_0$ relates inversely to $\xi$, fitting for $T_0$ is the usual method of measuring this quantity in the IQHE.

In the limit where frequency $f\approx \frac{k_B T}{h}$ hopping conduction should be dominated by the absorption of incident photons.  For temperatures below 1 Kelvin, these photons are in the microwave frequency range, $f \le 20$ GHz.  At $B=0$ Efros \cite{alefros} has calculated
\begin{equation}
Re\{\sigma_{xx}( \omega )\}= k  \epsilon_0 \epsilon \xi \omega \ \ \ \ \ \ {\rm and}\ k=\frac{\pi}{3}
\end{equation}
for hopping transport when Coulomb effects are included.  Polyakov and Shklovskii \cite{polyakov_shklovskii} have shown that Equation 2 holds in the IQHE for the limit $h f \ll k_B T$ but that in the zero tempature limit, $h f \gg k_B T$, $k=\frac{2 \pi}{3}$. Both results require that $\sigma_{xx}(\omega) \gg \sigma_{xx}^{DC}$, a condition which is satisfied across much of the IQHE plateaus.  Because of its simplicity, this result implies that measurements of $\sigma_{xx}$ across the plateau can be used to find $\xi$.  However, if strong Coulomb effects are absent, Mott\cite{mott1970} predicts $\sigma (\omega) \sim \omega^2$ for disordered hopping models independent of dimension.   Thus, measurement of $\sigma_{xx}(\omega)$ and determination through Equation 2 of $\xi$ coupled with comparison of $\xi$ measured using DC techniques and Equation 1 would be a useful consistency check of current hopping theory.

In this frequency range, we know of only 3 experiments \cite{engel,balaban,holhs1,holhs2} of which, Engel {\it et al.} \cite{engel} and Balaban {\it et al.} \cite{balaban} use configurations that give insufficient sensitivity when $\sigma_{xx}$ is small.  Very recently, Holhs {\it et al.} \cite{holhs2} have addressed the issues in Equation 2, finding $Re\{ \sigma_{xx}(\omega)\} \sim \omega$.

In this letter we present measurements for $\sigma_{xx}$ performed at frequencies 1.25 $\le f \le$ 10.05 GHz and for temperatures 235 mK $\le T \le$ 4.2 K.  Our results are consistent with Equation 2 and are in agreement with DC measurements of $\xi$ based on Equation 1.

	 The method we use is as follows.  The sample is mounted at one end of a coaxial resonator and by monitoring the amplitude of the reflected signal as $f$ is swept through resonance, we extract the the real and imaginary parts of $\sigma_{xx}$.  The resonator has a quality factor of approximately 1600 at helium temperatures and is fashioned from a piece of copper coaxial cable with a silver--plated center conductor.  The resonant frequencies are set by the length of the resonator and correspond to $f_n = {\rm n}\ \frac{2c}{\lambda}$.  This resonator geometry was chosen to allow the use of a Corbino disk pattern on the samples which creates a radial electric field.  Therefore, only radial currents in the sample, proportional to $\sigma_{xx}$, couple to the resonator, so $\sigma_{xx}$ is measured and $\sigma_{xy}$ is completely supressed. 
The resonator is kept undercoupled to the transmission line so to first order, the change in width of the resonant absorption, $\Delta W$, is directly proportional to the real part of $\sigma_{xx}$, i.e.
\begin{equation}
\Delta W = \frac{f_0}{2 \pi} \frac{ 4 Re(G_s) Z_0}{1+C_t/C_c} \propto Re(\sigma_{xx})
\end{equation}
Here $f_0$ is the fundamental frequency and $Z_0=50 \ \Omega$, the impedance of the resonator.  Our modeling of the impedances at the sample end of the resonator includes the conductance of the 2DEG, $G_s$, in parallel with the coupling capacitance, $C_c$, to the resonator, both in series with the capacitance of the GaAS substrate, $C_t$.   The value of $\Delta W$ and $f_{res}$ are found at each magnetic field point by using a four parameter least squares fit on the resonant absortion line data.  Henceforth all discussion of $\sigma_{xx}$ refers to the real part.

The resonator and sample are rigidly mounted in the tailstock of a dilution fridge in a superconducting solenoid.  The resonator is directly heat sunk to the mixing chamber. To provide local thermometry on the sample, a 1 k$\Omega$ RuO$_2$ resistor is mounted beside the sample and has its leads heat sunk to the sample holder.  Connection from the network analyzer to the resonator is through semirigid stainless steal coaxial cable into the cryostat which is  heat sunk at 4 K, the 1 K pot, the still, and the mixing chamber.  To stop thermal leakage down the center electrode a stripline patterned  onto a sapphire crystal heat sunk at 4 K is inserted.  Coupling between the transmission line and the resonator is tuned by the gap between the respective center electrodes.

 	The samples are standard GaAs/AlGaAs heterostructures.  The data we present here where measured using a sample with density $n=3.0 \times 10^{11}  {\rm \ cm^{-2}}$ and mobility $\mu=500,000 {\rm \ cm^2 \ V^{-1} \ s^{-1}}$ which we refer to as Sample 1 and a sample with $n=2.5 \times 10^{11} {\rm \ cm^{-2}}$ and mobility $\mu=100,000 {\rm \ cm^2 \ V^{-1} \ s^{-1}}$ (Sample 2).  In additon, a sample with $n=1.6 \times 10^{11}  {\rm \ cm^{-2}}$ and mobility $\mu=50,000 {\rm \ cm^2 \ V^{-1} \ s^{-1}}$ (Sample 3) was tested.  Photolithography was used to define metal contacts which were then evaporated on. These  consist of about 100$\AA$ Ni, 300$\AA$ Ge, and 600 $\AA$ Au.  Because of the capacitive nature of the contacts, annealing was not necessary.  Instead, a thin ($\sim 1\mu $m) polyimide \cite{polyimide} layer was spun onto the surface.  Once cured, this layer provides a very thin, tough layer which the resonator can be pressed up against without damaging the contacts or the 2DEG. 
 
\begin{figure}[tb]
\begin{center}
\leavevmode \epsfxsize \columnwidth
\epsffile{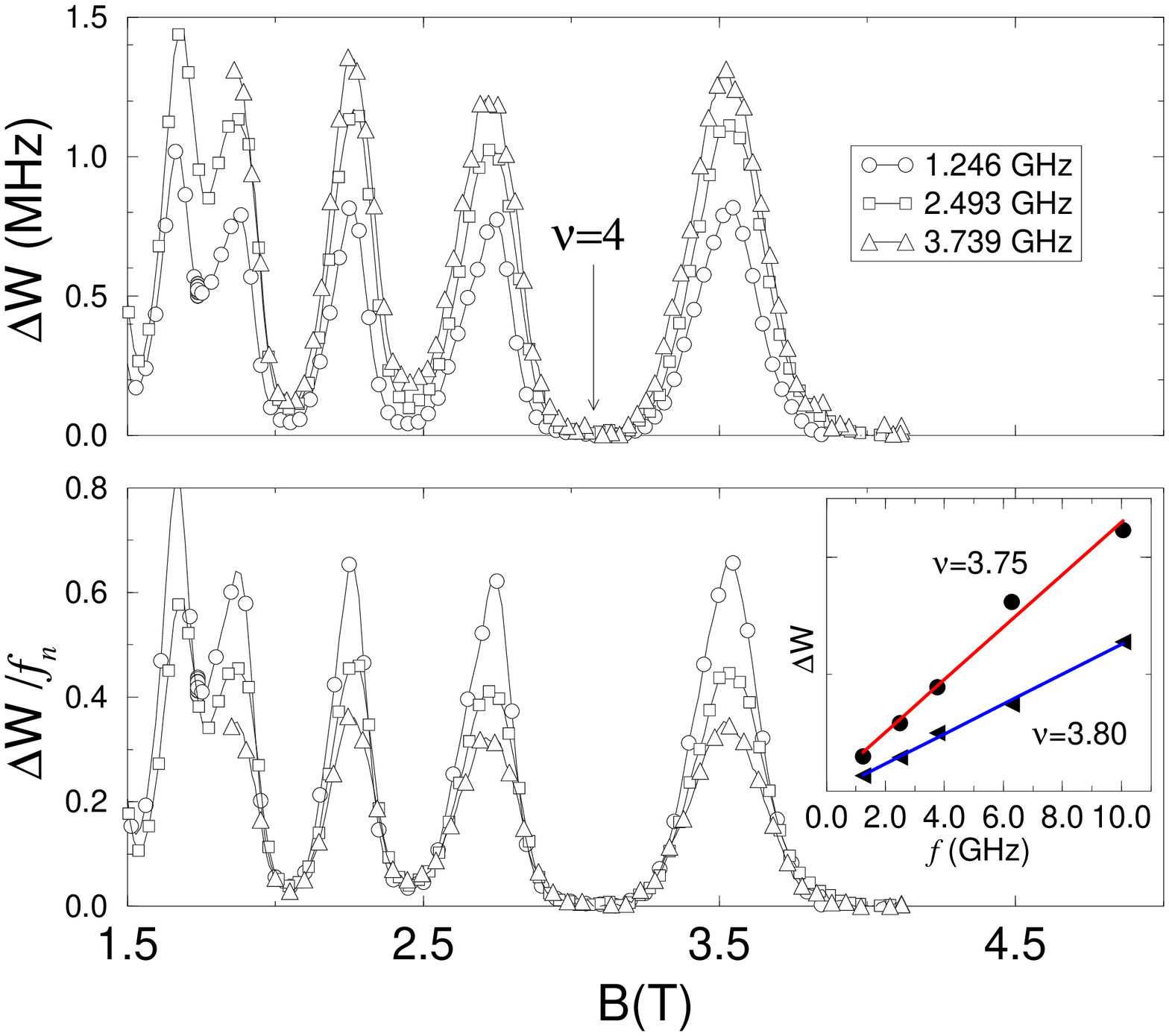}
\caption{{\bf Upper \ Panel}: $\Delta W$ vs. B  from Sample 1 taken at $f=$ 1.246, 2.493, and 3.739 GHz with $T=235$ mK. The data are shifted so that $\Delta W =0$ at $\nu=4$.  $\Delta W \sim \sigma_{xx}$. Data at different $f$ are distinct.  {\bf Lower\ Panel}: $\Delta W/f_n$ vs. B for the same 3 frequencies. The data collapse onto a single curve in accordance with Equation 2.  {\bf Inset\ panel}: $\Delta W$ vs. $f$ for two different fixed B points on the $\nu=4$ plateau demonstrating linear behavior over $1.246 \le f \le 10.05$ GHz.}
\label{fig.1}
\end{center}
\end{figure}

	The upper panel of Figure 1 plots the change in width of the resonant absorption peaks $\Delta W$ versus B for Sample 1 at the three lowest modes of our resonator, 1.246, 2.493, and 3.739 GHz.  Temperature is fixed at 235 mK and determined by the RuO$_2$ thermometer.  To avoid obscuring the data, only one third of the points in each set are plotted.  The unloaded line width of each resonance has been subtracted off each curve assuming that $\sigma_{xx}\approx0$ at $\nu=4$.  These data result from fitting either 80 points (1.246, 2.493 GHz) or 160 points (3.739 GHz) distributed over the resonant absorption peaks.  We checked that no heating of the 2DEG due to the incident microwaves was taking place by systematically decreasing the power until such effects vanished.  The maximum applied power absorbed by the sample is $- 66$ dBm.  Data taken using the same technique, but with a GaAs wafer lacking a 2DEG show no repeatable field dependence.
Note that data at different frequencies are visibly separate every where except the center of the $\nu=4$ plateau. At the other minima, the data show frequency dependence also.
	
	The lower panel of Figure 1 plots the same three data sets scaled by the appropriate frequency, i.e. $\Delta W/f_n$.  Again, only a fraction of the points are plotted with symbols.  Around the $\nu=4$ minimum the three data sets collapse onto a single curve for a wide range of B. Elsewhere, the $\sigma_{xx}$ minima at $\nu=5 \ {\rm and} \ 6$ at lower B, and the $\nu=3$ minimum at higher B also begin to collapse though not as thoroughly.  Scaling works best where E$_g$ is large, i.e. for small even $\nu$.  The $\sigma_{xx}$ peaks do not scale, but there, $\sigma_{xx} \approx \sigma_{xx}^{DC}$ and electrons are no longer localized.  Our data indicate scaling up to 10.05 GHz, the highest frequency at which we have made reliable measurements.  This is demonstrated in the inset to the lower panel where the $\Delta W$ versus $f$ is plotted at two different points on the $\nu=4$ plateau.  The lines are linear least squares fits to the data points and the slope is given by Equation 2.

\begin{figure}[tb!]
\begin{center}
\leavevmode \epsfxsize \columnwidth
\epsffile{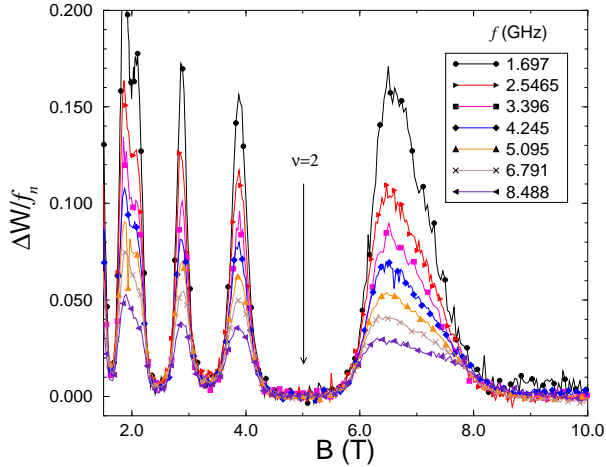}
\caption{$\Delta W$ vs. B measured at $1.7 \le f \le 8.5$ GHz on Sample 2 at T$=500$ mK. Clear frequency scaling is seen at $\nu=1 {\rm \ and \ } 2$.}
\label{fig.2}
\end{center}
\end{figure}

	In Figure 2, we plot $\Delta W/f_n$ measured using Sample 2 at a temperature of 500 mK over the frequency range $1.7 \le f \le 8.5$ GHz.  These data were measured in a resonator with $f_0 =0.85$ GHz and used higher measurement power levels but are similar in all other respects to data presented in Figure 1.  Convincing frequency scaling is seen around $\nu=2$ and with the exception of $f=1.697$ GHz, the data also collapse near the $\nu=1$ minimum.  Near $\nu=3$ incomplete frequency scaling is seen, and around $\nu=4$ the data scale for $f\ge 3$ GHz.  Data measured with Sample 3 at $T=500$ mK, which has roughly half the density of Sample 1, do not show scaling at this temperature.
  
\begin{figure}[tb]
\begin{center}
\leavevmode \epsfxsize \columnwidth
\epsffile{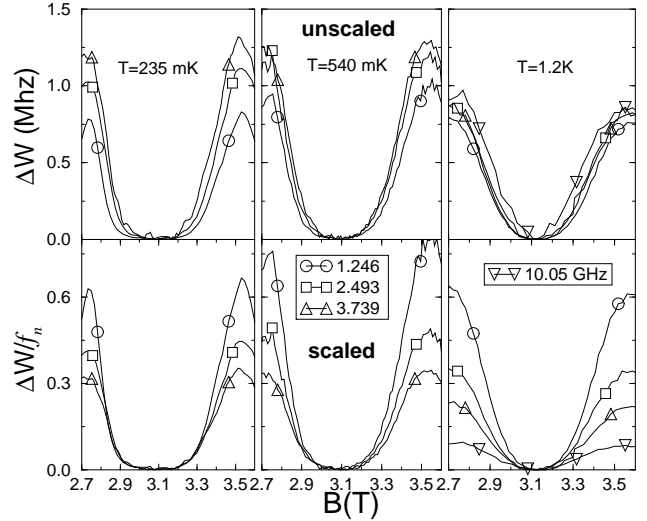}
\caption{{\bf Top Three Panels}: $\Delta W$ vs. B from Sample 1 at the  $\nu=4$ minimum presented for T=0.235, 0.540, and 1.2 K. The data are unscaled.  At T=235 mK each frequency is distinct.  By T=1.2 K this distinction vanishes.
{\bf Lower Three Panels}: $\Delta W/f_n$ vs. B at the same three temperatures,i.e. scaled data.  At 1.2 K data for $f=10.05$  GHz are included to demonstrate the absence of scaling over the $1.246 \le f \le 10.05$ GHz. }
\label{fig.3}
\end{center}
\end{figure}

In Figure 3 we present data for $\Delta W$ versus B around $\nu=4$ measured with Sample 1, both before scaling (upper windows) and after (lower) at temperatures of 235 mK, 540 mK, and 1.2 K.  The traces are for the three frequencies, 1.246, 2.493, and 3.739 GHz.  At 1.2 K, data for 10.05 GHz are included. Comparing the scaled data at 235 mK, 540 mK, and 1.2 K it is evident that the scaling region around $\nu=4$ grows narrower as temperature is increased and vanishes altogether by 1.2 K.  Data taken at 4.2 K for $1.7 \le f \le 6.6$ GHz do not show any return of scaling.  Nor does data at $\nu=2$ scale at 4.2 K despite the hopping region being proportionately wider.  This is born out by looking at the unscaled data which become more similar as the temperature is increased and frequency effects diminish in relative importance.  In the two lower temperature plots, the three frequencies are clearly separated.  At a temperature of 1.2 K the three lowest frequency traces are nearly indistinguisable and the 10.05 GHz data varies little from the other three frequencies.  At 4.2 K, the data seem independent of frequency over this range.

The data presented so far demonstrate that in the appropriate limits, i.e. when $h f \ge k_BT$ and $\sigma_{xx}(\omega) \gg \sigma_{xx}^{DC}$ \cite{limits} that $\sigma_{xx}(\omega) \sim \omega$.  This linear dependence implies that Coulomb effects are important in AC hopping conduction in the IQHE.  As $T$ increases, the range of B over which frequency scaling is seen decreases and eventually vanishes, even for large $f$, by $T=1.2$ K. This is in agreement with Reference \cite{polyakov_shklovskii}.  The size of E$_g$ between extended states affects when scaling is seen, since frequency scaling appears at higher $T$ for even plateaus than for spin--split plateaus.
	It is implied in Reference \cite{polyakov_shklovskii} that a crossover between $h f \gg k_B T$ and $h f \ll k_B T$ should be seen in the frequency scaling. We do not see such a crossover in measurements from $T=235$ mK up to $T=4.2$ K.  

\begin{figure}[tb]
\begin{center}
\leavevmode \epsfxsize \columnwidth
\epsffile{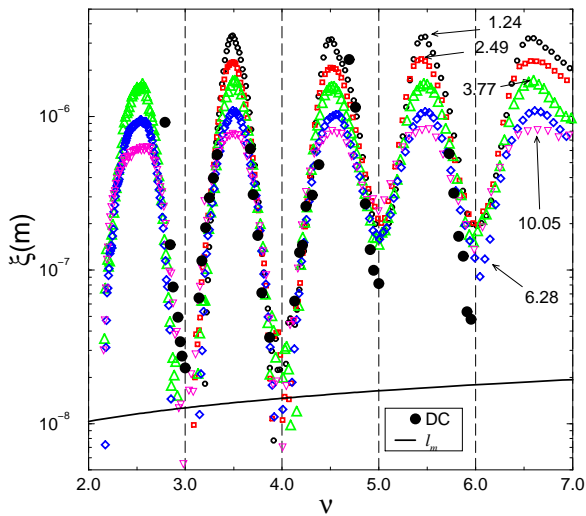}
\caption{$\xi$ vs. $\nu$.  The open symbols derive from $\Delta W$ versus B data at $f=1.246,\ 2.493,\ 3.770,\ 6.282,{\rm \ and \ } 10.05$ GHz taken with Sample 1.  The filled  circles are DC measurements of $\xi$ from a peice of the same wafer.  The solid line is the magnetic length. Where scaling is good, the AC and DC measurements of $\xi$ agree.}
\label{fig.4}
\end{center}
\end{figure}

In Figure 4, we plot $\xi$ versus filling factor $\nu$.  The open symbols derive from measurements of $\Delta W$ versus B on Sample 1 at $1.246 \le f \le 10.05$ GHz at $T=235$ mK.  Combining Equations 1 and 3 with a knowledge of the geometry of our sample, we arrive at the expression,
\begin{equation}
\xi= \frac{\Delta W}{f} \ (1+\frac{C_t}{C_c}) (1.5 \times 10^{-4}{\rm \ m})
\end{equation}  
The factor $1 + C_t/C_c$ is found to be about 17 if a 20 $\mu$m gap is used in the coupling capacitor.  Imprecise knowledge of this factor limits the precision in vertical placement of the data.  Scatter at the bottom of the minima are represenative of noise levels.  The filled circles represent measurements of $\xi$ at DC with a piece of the same wafer.  The data are calculated by fitting $\sigma_{xx}$ vs. $T$ data to Equation 1 and extracting $T_0$.  The solid line is the magnetic length of the system given by $l_m=(\frac{eB}{h})^{\frac{1}{2}}$ and as noted, the density of Sample 1 is $n=3.0 \times 10^{11} {\rm \ cm^{-2}}$.

	Comparison of the DC $\xi$ with the high frequency data show qualitative agreement in the high B plateaus, $\nu=3 {\rm \ and \ }4$ where the frequency scaling prevails but do not agree at $\nu=5 {\rm \ and \ } 6$ where the scaling is incomplete.  Near half integer $\nu$ the high frequency data disagree with DC data points but since $\sigma_{xx}(\omega) \approx \sigma_{xx}^{DC}$ the theory is not expected to work in the peaks.

	Although DC hopping conduction is observed to obey Equation 1, several factors could lead to the e$^{-(\frac{T_0}{T})^{\frac{1}{2}}}$ factor observed in experiments, for example, variable range hopping of weakly interacting electrons along disordered 1--D channels.  The two measurements of $\xi$, high frequency and the DC data, are both consistent with hopping models that include the Coulomb gap.  The agreement of these data with each other points to the correctness of Coulomb gap arguements in dealing with hopping conduction when interactions are strong.

In summary, we have measured $\sigma_{xx}(\omega)$ for 2DEG's and find that the conductivity scales linearly with frequency in two different samples provide that $E_g \gg k_B T \ge h f$.  The B region over which we observe scaling diminishes smoothly as $T$ increases.  The temperature at which scaling arrives is dependent on E$_g$.  Above 1 K, $\sigma_{xx}$ shows no frequency dependence up to 10 GHz.  The localization lengths determined from our high frequency data agree with data from the same wafer taken at DC.  These measurements support most aspects of the theory by Polyakov and Shklovskii \cite{polyakov_shklovskii} and indicate the importance of Coulomb gap effects in 2-D hopping.

The authors gratefully acknowledge stimulating discussions with H.\ P.\ Wei and Steven M.\ Girvin.  Samples were grown by J.\ E.\ Cunningham at Bell Labs and K.\ Alavi at U.\ Texas.


\begin{thebibliography}{}
\bibitem{limits}Implicit here is the assumption that frequency is still low enough that coulomb effects are important and that the IQHE has not be washed out by high frequency.
\end{thebibliography}
\end{document}